\documentclass[12pt,aps,showpacs]{revtex4}
\usepackage{epsfig,amsmath,amssymb,amsfonts}
\usepackage{bbm}

\newcounter{fig}

\newcommand{\nc}{\newcommand}
\nc{\be}{\begin{equation}}
\nc{\ee}{\end{equation}}
\nc{\bea}{\begin{eqnarray}}
\nc{\eea}{\end{eqnarray}}
\nc{\bi}[1]{\bibitem{#1}}
\nc{\lsim}{\mbox{\raisebox{-.6ex}{~$\stackrel{<}{\sim}$~}}}
\nc{\gsim}{\mbox{\raisebox{-.6ex}{~$\stackrel{>}{\sim}$~}}}

\begin{document}

\rightline{HD-THEP-04-11}

\vskip 0.2in

\title{\Large Unruh response functions for scalar fields in de Sitter space}

\author{\Large Bj\"orn Garbrecht$^{(1)*}$, Tomislav Prokopec$^{{(1)(2)}}$
       }

\email[]{B.Garbrecht@thphys.uni-heidelberg.de}
\email[]{T.Prokopec@thphys.uni-heidelberg.de}
%\email[]{M.G.Schmidt@thphys.uni-heidelberg.de}

\bigskip

\affiliation{$^{(1)}$Institut f\"ur Theoretische Physik, Heidelberg University,
             Philosophenweg 16, D-69120 Heidelberg, Germany}

\affiliation {$^{(2)}$Max-Planck-Institut f\"ur Gravitationsphysik
                       (Albert-Einstein-Institut)
             Am M\"uhlenberg 1, D-14476 Golm, Germany}

\date{\today}

\begin{abstract}

We calculate the response functions of a freely falling
Unruh detector in de Sitter space coupled to scalar fields of different coupling
to the curvature, including the minimally coupled massless case.
Although the responses differ strongly in the infrared as a consequence of
the amplification of superhorizon modes, the energy levels of the
detector are thermally populated.

\end{abstract}

\pacs{98.80.Cq, 04.62.+v, 98.80.-k}
%98.80.-k   Cosmology
%04.62.+v   Quantum field theory in curved spacetime
%98.80.Cq   Particle-theory and field-theory models of the early Universe
%           (including cosmic pancakes, cosmic strings, chaotic phenomena,
%           inflationary universe, etc.)

\maketitle

%
%%%%%%%%%%%%%%%%%%%%%%%%%%%%%%%%%%%%%%%%%%%%%%%%%%%%%%%%%%%%%%%%%%%%%%%%%%%%%%%
%  MAIN TEXT
%%%%%%%%%%%%%%%%%%%%%%%%%%%%%%%%%%%%%%%%%%%%%%%%%%%%%%%%%%%%%%%%%%%%%%%%%%%%%%%
%

\section{Introduction}

It is expected that an observer, corresponding to an
Unruh detector~\cite{Unruh:1976,BirrellDavies:1984} coupled to a scalar field,
when freely falling in de Sitter space, will perceive radiation with a thermal
spectrum of the de Sitter temperature
$T_H = H/(2\pi)$~\cite{GibbonsHawking:1977,BirrellDavies:1984},
where $H$ denotes the Hubble parameter. It is the purpose of this
article, to clarify in what sense this result is universal to scalar
fields of different couplings to the de Sitter background and how the detector
apprehends the differences.

Let us therefore refine what is meant by the observation of thermal radiation:
At first order
in perturbation theory, the detector response function is proportional to
the Fourier transform of the scalar propagator {\it w.r.t.} the proper
time of the detector, and it describes how many particles are absorbed and
emitted per unit time.
When being
in equilibrium with the de Sitter background, the energy levels of 
the detector are thermally populated according to the temperature
$T_H$.

The easiest way to derive this is to note that the propagator
for nonminimally coupled scalars
has in the imaginary direction of its proper time $t$ the periodicity
$t\rightarrow t+2\pi i/ T_H$~\cite{GibbonsHawking:1977,BoussoMaloneyStrominger:2001,SpradlinStromingerVolovich:2001}.
As we will point out,
this however does not completely characterize the response function of
the detector, which describes the number of particles detected per unit time.
The rate turns out to depend on the scalar mass and on its coupling to
the curvature, as was first shown in Ref.~\cite{Higuchi:1986}, circumventing
the use of the scalar propagator.

The fact that in de Sitter space the invariance of the quantum vacuum becomes
manifest when the scalar propagator only depends
on the proper time seperation along a geodesic has led to the practice of
defining the de Sitter vacua through this
quantity~\cite{ChernikovTagirov:1968,BunchDavies:1978,Mottola:1984,Allen:1985,AllenFolacci:1987}.
However, in the case of
a massless scalar, which is minimally coupled to the curvature, this leads to
a problem since the propagator is infrared divergent~\cite{Allen:1985}.
We argue that, when regulated by a cutoff, this divergence gives rise to
a contribution which is irrelevant to the total detector response.
In addition, we compare with the response functions of a detector
immersed in a thermal bath in Minkowski space and discuss the
situation in de Sitter space-times with dimension other than four.

\section{Unruh's detector}

Unruh's 
detector~\cite{Unruh:1976,BirrellDavies:1984,SpradlinStromingerVolovich:2001}
corresponds to a heavy slowly moving particle along a trajectory
$x=x(t)$, which interacts with a scalar bath of particles as
\begin{equation}
 {\cal L}_{\rm Unruh} = - h \hat m(t) \Phi\big(x(t)\big)
\label{Unruh detector}
\,,
\end{equation}
where $h$ is a coupling constant, $t$ is the proper time
and $\hat m(t)$ represents the quantum operator describing the monopole
interaction of the detector with the scalar bath $\Phi=\Phi(x)$. 
Since the detector is assumed to be very heavy, it does not fluctuate 
in space, and hence the only time dependence is in $\hat m(t)$. 
The response of Unruh's detector can be derived as follows. 
One assumes that the state of the detector is specified by
a set of energy eigenstates, $\{|E\rangle \}$, and that each absorbtion
of a scalar quantum elevates the energy of the detector by
$\Delta E = E-E_0$; the converse is true for each emission.

Consider now at first order perturbation theory the transition amplitude
${\cal M}$ from a state
$|E_0\rangle\otimes|\varphi_0\rangle$ to a state 
$\langle E|\otimes\langle\varphi|$
\begin{equation}
{\cal M}(E_0\rightarrow E;t_0,t_f) =  
   h m_{E_0,E} \int_{t_0}^{t_f} dt_1\, {\rm e}^{-i(E-E_0)t_1} 
           \langle\varphi|\Phi\big(x(t_1)\big)|\varphi_0\rangle
\,,
\label{transition amplitude}
\end{equation}
where $m_{E_0,E}
   \equiv \langle E|\hat m(0)|E_0\rangle$ 
($\hat m(t_1) = {\rm e}^{i\hat Ht_1}\hat m(0){\rm e}^{-i\hat Ht_1}$)
is defined such that it does not include stimulated emission.
In order to take account of the influence of initial conditions, 
for simplicity we assume that the interaction $\hat m(t_1)$ switches on
at $t_1 = t_0$ as $\Theta(t_1-t_0)$, where $\Theta(x) = 1$ when $x>0$,
and $\Theta(x) = 0$ when $x<0$. More generally, one would expect
that the interaction switches on adiabatically. Since we are not interested in
studying in detail how the response function depends on 
the initial conditions and on how the interaction turns on,
this rough treatment should suffice.
Upon
%taking the limit $t_0\rightarrow -\infty$, 
squaring ${\cal M}$ and making use of orthonormality
and completeness of the scalar eigenstates $\{|\varphi\rangle\}$, 
and taking $t_0\rightarrow - \infty$, 
we get for the transition probability per unit proper time, 
\begin{equation}
  \frac{d P(E_0\rightarrow E)}{dt} 
       = h^2 |m_{E_0,E}|^2\; \frac{d{\cal F}(\Delta E)}{dt}
\qquad (\Delta E = E  - E_0)
\,,
\label{transition probability}
\end{equation}
where
\begin{equation}
\frac{d{\cal F}(\Delta E)}{dt}  
          = \int_{-\infty}^{\infty} d \Delta t 
             \,{\rm e}^{-i\Delta E\Delta t}
                    iG^{<}\Big(x(t+\Delta t/2);x(t-\Delta t/2)\Big)
\,
\label{response function}
\end{equation}
is the response function per unit proper time.
Here $iG^{<}(x_1;x_2) 
= \langle \varphi_0|\Phi(x_2)\Phi(x_1)|\varphi_0\rangle $ is the
(positive frequency) coordinate space Wightman function,
%In Eqs.~(\ref{transition probability}) and (\ref{response function})
and we defined the time variables, $t\equiv(t_1+t_2)/2$
and $\Delta t \equiv t_1 - t_2$.
The expression~(\ref{response function}) is real by construction,
which can be formally shown by taking the complex conjugate of the integrand
in~(\ref{response function}) and then making use of the hermiticity property
of the Wightman functions, $[iG^{<,>}(x_1,x_2)]^* = iG^{<,>}(x_2,x_1)$.
We shall study the particle spectrum observed in an appropriate vacuum state
$| \varphi_0\rangle=|0\rangle$, the precise nature of which will be specified
later. It is generally accepted that it is most natural to 
consider (a freely-falling) Unruh detector moving on a geodesic,
since in this case the detector does not see additional particles due to the
Unruh effect, which would be present only because of the  acceleration
of the detector. 

 A nice property of Unruh's detector is
the  separability of the transition probability~(\ref{transition probability}) 
into a product of the selectivity function, 
which depends on the inner structure of the detector, which is 
completely specified by the operator $\hat m(t)$ and the coupling $h$,
and the response function, which depends on the state of the 
scalar field only. It appears useful to rewrite
Eq.~(\ref{response function}) in terms of the Wigner function
as 
\begin{equation}
  \frac{d{\cal F}}{dt}
            = \int \frac{d^3 k}{(2\pi)^3} 
              iG^<\big(k^0 = \Delta E,\vec k,x(t)\big)\,,
\label{response function:2}
\end{equation}
where the Wigner function is defined as the Fourier transform {\it w.r.t.}
the relative coordiante of the Wightman  function, 
\begin{equation}
  iG^<(k,x) = \int d^4 r {\rm e}^{ik\cdot r} iG^<(x+r/2, x-r/2)
\,,
\label{Wigner function}
\end{equation}
such that $r^0 = x_1^0 - x_2^0$ are proper times. 
This means that the response function of Unruh's detector 
is completely insensitive to particle momenta (which is consistent with the 
assumption that the detector must be very massive), and it measures 
(absorbs) scalar particles of all possible momenta $\vec k$; likewise,
it isotropically emits particles of all momenta.

\section{Thermal response}

In order to get an insight in how a certain distribution function
of a scalar field is perceived by the detector, we consider the response
function for the Bose-Einstein distribution.
The Wightman function for a thermally excited scalar field
has the well known form~\cite{LeBellac:1996}
\begin{eqnarray}
   iG_{\tt th}^<(k)
      = 2\pi\textnormal{sign}(k^{0})\delta(k^{2}+m_\phi^{2})
        \frac{1}{{\rm e}^{\beta k^{0}}-1}
\,,
\label{G:thermal}
\end{eqnarray}
where $\beta = 1/T$ denotes the inverse temperature, 
and $m_\phi$ the scalar mass.
The response function~(\ref{response function:2}) for $d$ dimensions in
flat space-time is easily calculated, 
\begin{equation}
 \frac{d{\cal F}_{{\tt th},d}(\Delta E)}{dt} 
   = \frac{2^{2-d}\pi^{\frac{3-d}{2}}}{\Gamma(\frac{d-1}{2})}
{\rm sign}(\Delta E)\Theta\left((\Delta E)^2 - m_\phi^2\right)
                  \left((\Delta E)^2 - m_\phi^2\right)^{\frac{d-3}{2}}
    \,\frac{1}{{\rm e}^{\beta \Delta E} -1}
\,,
\label{G:thermal:ddim}
\end{equation}
and explicitly, for $d=4$, one finds
\begin{equation}
 \frac{d{\cal F}_{{\tt th},d=4}(\Delta E)}{dt} 
   = \frac{{\rm sign}(\Delta E)\Theta\left((\Delta E)^2 - m_\phi^2\right)
                  \sqrt{(\Delta E)^2 - m_\phi^2}}
          {2\pi}
    \,\frac{1}{{\rm e}^{\beta \Delta E} -1}
\,.
\label{G:thermal:2}
\end{equation}
Hence, the response function of Unruh's detector to a 
scalar thermal state contains, apart from the Bose-Einstein distribution,
an additional factor, which depends on the scalar particle mass and
reduces to $\Delta E/(2\pi)$ in the massless limit and $d=4$.
This is precisely the factor that arises for the conformal scalar 
vacuum~(\ref{response function:conformal:2}) in 
de Sitter space~\cite{BirrellDavies:1984}.

\vskip 0.1in

\section{Scalar fields in de Sitter inflation}

 The Lagrangean for a massive real scalar field in
a curved space-time background is given by 
\begin{equation}
\sqrt{-g}\mathcal{L}_\Phi = - \frac{1}{2}\sqrt{-g}g^{\mu\nu}
                       (\partial_{\mu}\Phi)(\partial_{\nu}\Phi)
                     - \frac{1}{2}\sqrt{-g}(m_\phi^{2}+ \xi {\cal R})\Phi^{2}
,
\label{Lagrangian:scalar}
\end{equation}
where $g = {\rm det}[g_{\mu\nu}]$ denotes the determinant of the metric
$g_{\mu\nu}$,  $g^{\mu\nu}$ is the inverse of the metric, 
$m_\phi$ the scalar mass and ${\cal R}$ the curvature scalar.
In conformal space-times, in which the metric is of the form
\begin{equation}
 g_{\mu\nu} = a^2 \eta_{\mu\nu}
\,, 
\label{conformal metric}
\end{equation}
where $\eta_{\mu\nu} = {\rm diag}(-1,1,1,1)$ is the flat Minkowski metric
and $a=a(\eta)$ the scale factor, 
the Lagrangean~(\ref{Lagrangian:scalar}) reduces to
\begin{equation}
\sqrt{-g}\mathcal{L}_\Phi = - \frac{1}{2}a^{2}\eta^{\mu\nu}
                       (\partial_{\mu}\Phi)(\partial_{\nu}\Phi)
                     - \frac{1}{2}a^4(m_\phi^2 + \xi {\cal R})\Phi^2
\,.
\label{Lagrangian:scalar:conformal}
\end{equation}
For example, in (a locally) de Sitter inflation
$a = -1/(H\eta)$ ($\eta<0$), ${\cal R} = 12 H^2$ (in $D=4$ space-time 
dimensions), while in radiation-matter era, 
$a = a_r\eta+a_m\eta^2$, where $H$ denotes the Hubble parameter, 
$\eta$ conformal time, $a_r$ and $a_m$ are constants. 

 The special cases of interest are:

\begin{itemize}
\item
conformally coupled massless scalar field
($\xi = 1/6$, $m_\phi = 0$), for which a conformally rescaled 
scalar $\varphi \equiv a\Phi$ satisfies the simple differential equation 
($\nabla$ denotes a spatial derivative)
\begin{equation}
 (\partial_\eta^2 - \nabla^2)\varphi(x) = 0
\,;
\label{conformally coupled scalar}
\end{equation}

\item
and nearly minimally coupled light scalar 
($|\xi| \ll 1 $, $m_\phi \ll H$),  which obeys 
\begin{equation}
 \Big(
      \partial_\eta^2 - \nabla^2 - \frac{1}{a}\frac{d^2 a}{d\eta^2}
    + a^2(m_\phi^2 + \xi {\cal R})
 \Big)\varphi(x) = 0
\,.
\label{nearly minimally coupled light scalar}
\end{equation}
(in de Sitter inflation, $-a^{-1}{d^2 a}/d\eta^2 = -2/\eta^2 = -2a^2 H^2$);

\item
and the minimally coupled massless scalar 
($\xi = 0$, $m_\phi = 0$),  which satisfies the following
differential equation, 
\begin{equation}
 \Big(
      \partial_\eta^2 - \nabla^2 - \frac{1}{a}\frac{d^2 a}{d\eta^2}
 \Big)\varphi(x) = 0
\,.
\label{minimally coupled scalar}
\end{equation}

\end{itemize}

\subsection{Conformal vacuum in de Sitter space}

 We now calculate the response function~(\ref{response function})
for a conformally coupled massless scalar 
field~(\ref{conformally coupled scalar}) ($\xi = 1/6$, $m_\phi = 0$), 
for which the de Sitter invariant Green function during inflation
in $D=4$ reads~\cite{BirrellDavies:1984}
\begin{equation}
 i G_{\rm conf}(y)  =   \frac{H^2}{4\pi^2} \frac{1}{y}
\,,
\label{Delta:conformally coupled scalar}
\end{equation}
where $y$ denotes the de Sitter length function
\begin{equation}
   y = \frac{\Delta x^2}{\eta_1\eta_2}
     \equiv 4 \sin^2\Big(\frac 12 H\ell \Big)
\label{y:+-},
\end{equation}
which is related to the geodesic distance $\ell$ as indicated,
and $\Delta x^2=-(\eta_1-\eta_2)^2 +\|\vec x_1 - \vec x_2\|^2$.
For an observer moving along a geodesic, 
$y=y\big(x(t+\Delta t/2);x(t-\Delta t/2)\big) 
 = - 4 \sinh^2\big(H\Delta t/2 \big)$.

\begin{figure}[tbp]
\vskip -0.3in
\centerline{\hspace{.in} 
\epsfig{file=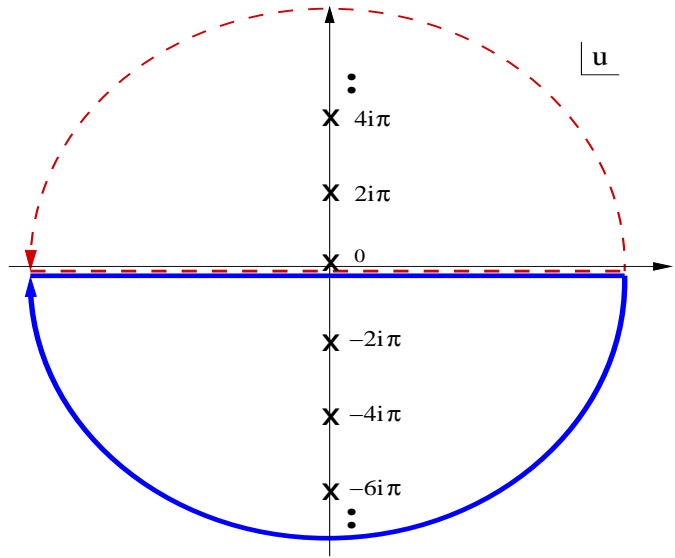, width=3.5in,height=2.9in}
}
\vskip -0.1in
\caption{\small
  The integration contour for the Unruh's
  detector response function in conformal vacuum.
The solid (blue) contour corresponds to $\Delta E > 0$; the dashed (red) 
contour to $\Delta E < 0$.
}
\label{fig-1}
\end{figure}

The response function~(\ref{response function})
for the conformal vacuum~(\ref{Delta:conformally coupled scalar}) 
can then be written as
\begin{equation}
   \frac{d{\cal F}_{\rm conf}(\Delta E)}{d t}
          = -\frac{H}{4\pi^2}\int_{-\infty}^{\infty} du
             \,{\rm e}^{-i\Delta E u/H}
       \frac{1}{4\big[\sinh\big(u/2\big) - i\epsilon\big]^2}
\,\,,
\label{response function:conformal}
\end{equation}
where the pole prescription corresponds to that of the 
Wightman function $iG^<$.
This integral can be easily performed by contour integration. The (double)
poles (which also correspond to the {\it zero}s of $y$) all lie on the
imaginary axis, $u_n = Ht_n = 2i\pi n$ $(n\,\epsilon\,\mathbbm{Z})$. 
For $E>E_0$ 
the contour of integration ought to be closed by a large circle below the real
axis, such that the integral in~(\ref{response function:conformal}) 
can be evaluated by summing the residua which lie (strictly)
below the real axis, as illustrated in figure~\ref{fig-1}. The result is 
\begin{eqnarray}
   \frac{d{\cal F}_{\rm conf}(\Delta E)}{d t}
          &=& -\frac{H}{4\pi^2}(-2\pi i) \sum_{n=-1}^{-\infty} 
                \frac{-i\Delta E}{H} {\rm e}^{(2\pi \Delta E/H)n}
\nonumber\\           
          &=& \frac{\Delta E}{2\pi} \, 
              \frac{1}{{\rm e}^{(2\pi/H)\Delta E}-1}
\,,
\label{response function:conformal:2}
\end{eqnarray}
which is identical to the response function~(\ref{G:thermal:2}) 
of the thermal Bose-Einstein distribution for massless scalars,
confirming thus the well known result~\cite{BirrellDavies:1984}.
For $\Delta E <0$, the contour should be closed above the real axis,
such that the contributing poles are $n\geq 0$, also shown in 
figure~\ref{fig-1}.  The result of integration 
is again given by Eq.~(\ref{response function:conformal:2}).

\subsection{Nearly minimally coupled light scalar}
\label{Nearly minimally coupled light scalar}

 The Green function for a massive scalar field coupled to gravity 
as indicated by the Lagrangean~(\ref{Lagrangian:scalar:conformal})
is given by the 
Chernikov-Tagirov~\cite{ChernikovTagirov:1968}
(Bunch-Davies~\cite{BunchDavies:1978}) vacuum 

\begin{equation}   
 iG(y) =  \frac{H^2}{4\pi^2}\Gamma\Big(\frac 32 - \nu\Big)
                             \Gamma\Big(\frac 32 + \nu\Big)
             \; {}_2 F_1 \Big(\frac 32 - \nu,
                              \frac 32 + \nu, 
                              2;
                              1-\frac y4
                          \Big)
\,,
\label{ChernikovTagirov}
\end{equation}
where 
\begin{equation}
 \nu = \sqrt{\Big(\frac{3}{2}\Big)^2 - \frac{m_\phi^2 + 12\xi H^2}{H^2}}
\,.
\label{nu}
\end{equation}
The uniqueness of $iG(y)$ follows from the requirement that 
the lightcone singularity is of the Hadamard form.

When expanded in powers of 
\begin{eqnarray}
      {\tt s} &\equiv& \frac{3}{2}
                 - \nu 
%                -    \bigg[
%                       \Big(
%                          \frac{3}{2}
%                       \Big)^2
%                         -  \frac{m_\phi^2 + \xi {\cal R}}{H^2}
%                      \bigg]^\frac12
%\nonumber\\
                =   \frac{m_\phi^2 }{3H^2} + 4\xi
                 +    O\Big(\big[(m_\phi^2/H^2) + 12\xi\big]^2\Big)
\,,\qquad |{\tt s}| \ll 1
\,,
\label{s:def}
\end{eqnarray}
the Green function for the Chernikov-Tagirov vacuum reduces to the
following simple form~\cite{ProkopecPuchwein:2003}
\begin{equation}
 iG(y;{\tt s})  = 
     \frac{H^2}{4\pi^2}\bigg\{
                              \frac{1}{y}
                            - \frac 12 \ln(y)
                            + \frac 1{2{\tt s}}
                            - 1 + \ln(2)
                            + O({\tt s})
                    \bigg\}
\,.
\label{iDelta:4dim:massive}
\end{equation}

The nontrivial new integral comes from the term
$ i G^<_{m = 0}\propto \ln(y)$, and its contribution to the response function
yields the integral
%\ref{footnote1}), 
%
\begin{equation}
 \frac{d{\cal F}_{\rm ln}(\Delta E)}{d t}
          = -\frac{H}{8\pi^2}\int_{-\infty}^{\infty} du
             \,{\rm e}^{-i \Delta E u/H}
               \bigg[
                     \ln\Big(4\sinh^2\big(u/2\big)\Big)
                 + i\pi {\rm sign}(u)
               \bigg]
\,,
\label{respone function:2}
\end{equation}
where we broke the logarithm into the real and imaginary contributions, 
in accordance with the $\varepsilon$-prescription for $iG^<$. 
The real part of the logarithm can be evaluated by
 breaking it into positive and negative $u$ 
and then performing a partial integration
(or, as it is more commonly done, by expanding the logarithm), while the
imaginary part can be integrated trivially:
\begin{eqnarray}
 \frac{d{\cal F}_{\rm ln}(\Delta E)}{d t}
&=& \frac{H^2}{4\pi^2\Delta E}\int_{0}^\infty du
   \sin\left(\frac{\Delta E}{H}u\right)\coth\left(\frac{u}{2}\right)
          - \frac{H^2}{4\pi\Delta E}\nonumber\\
 &=& \frac{H^2}{2\pi \Delta E}
                \frac{1}{{\rm e}^{2\pi\Delta E/H}-1}
\,,
\label{respone function:2a}
\end{eqnarray}
where in the last step we made use of Eq.~(3.981.8) of 
Ref.~\cite{GradshteynRyzhik:1965}.

 The remaining integrals in the response function~(\ref{response function})
simply yield $\delta$-function contributions, 
\begin{equation}
 \frac{d{\cal F}_{\delta}(\Delta E)}{d t}
          = \frac{H^2}{2\pi} \delta(\Delta E) 
\Big(\frac{1}{2 \tt s} - 1 + \ln(2) + O({\tt s})\Big)
\,.
\label{respone function:rest}
\end{equation}

Collecting all terms together, we get the response function for the nearly
minimally coupled massive scalar:
\begin{eqnarray}
\frac{d{\cal F}_{m_\phi \neq 0}(\Delta E)}{d t}
          &=&  \frac{\Delta E}{2\pi}\bigg(1+\frac{H^2}{\Delta E^2}\bigg)
                   \frac{1}{{\rm e}^{(2\pi/H)\Delta E}-1}
           +  \frac{H^2}{2\pi} \delta(\Delta E) 
%                 \Big(\frac{3H^2}{2(m_\phi^2 + 12 \xi H^2)} - 1 + \ln(2)\Big) 
                 \Big(\frac{1}{2{\tt s}} - 1 + \ln(2) +  O({\tt s})\Big)
%\nonumber\\
%\,,\qquad {\tt s} = \frac{m_\phi^2}{3H^2} + 3\xi\,, \qquad |{\tt s}\ll 1
\,.\qquad
\label{respone function:m}
\end{eqnarray}

\subsection{Minimally coupled massless scalar field}
\label{Minimally coupled massless scalar field}

The Green function of a minimally coupled massless scalar field exhibts an
infrared divergence~\cite{Allen:1985}, and the construction of a finite
propagator necessarily breaks de Sitter invariance. For the purpose of
calculating loop diagrams and using dimensional regularisation, one
considers the propagator in $d$ dimensions with appropriate
counterterms to cancel the infrared divergence. For a detailed discussion, see
Ref.~\cite{OnemliWoodard:2004}. In four dimensions, one
obtains~\cite{ProkopecWoodard:2003,ProkopecPuchwein:2003}
\begin{equation}
 i G_{m = 0}(x_1;x_2)  = 
     \frac{H^2}{4\pi^2}\bigg\{
                              \frac{1}{y}
                  - \frac 12 \ln(y) 
                  + \frac 12 \ln\big(a(\eta_1)a(\eta_2)\big)
                  -\frac 14
                            +\ln(2)
                    \bigg\}\,,
\label{iDelta:4dim:massless}
\end{equation}
where the term $\propto \ln\left(a(\eta_1)a(\eta_2)\right)$ breaks
de Sitter invariance.

Yet, this does not imply, that there is no de Sitter-invariant vacuum,
as pointed out in Ref.~\cite{KirstenGarriga:1993}, where such an invariant
state is explicitly constructed by quantizing the mode with zero momentum
seperately. However, singling out that mode does not render the propagator
finite, as one sees when regulating the propagator with an infrared cutoff 
$k^0$, such that one obtains~\cite{TsamisWoodard:1993}
\begin{equation}
 i G_{m = 0,k^0}(x_1;x_2)  = 
     \frac{H^2}{4\pi^2}\bigg\{
                              \frac{1}{y}
                  - \frac 12 \ln(y)
                     + \frac 12 \ln\big(a(\eta_1)a(\eta_2)\big)
                            -\ln(k^0 H) -\gamma_E + O(k^0)
                    \bigg\}
\label{propagator:regulated}
\,,
\end{equation}
where $\gamma_E = 0.577215..$ is Euler's constant.
This expression differs from the propagator~(\ref{iDelta:4dim:massless}),
which we shall use for calculating the response, only by a constant.

The response function is easily reconstructed from the results of
section~\ref{Nearly minimally coupled light scalar}:
\begin{eqnarray}
\frac{d{\cal F}_{m_\phi = 0}(\Delta E)}{d t}
           &=&  \frac{\Delta E}{2\pi}\bigg(1+\frac{H^2}{\Delta E^2}\bigg)
                      \frac{1}{{\rm e}^{2\pi\Delta E/H}-1} 
            +  \frac{H^2}{2\pi} \delta(\Delta E) 
                 \Big(\ln(a) -\frac 14 + \ln(2) \Big) 
\,,
\label{respone function:massless}
\end{eqnarray}
where $\ln(a) = Ht = N$ is the number of e-folds elapsed since the 
beginning of inflation, if we set the initial scale factor to be one.
This contribution to the response function vanishes for all $\Delta E\not= 0$,
such that under the assumption that its energy levels are 
not degenerate, the detector is insensitive to the breaking of de Sitter
invariance by the propagator.
Note also that, when integrated over $\Delta E$ around $\Delta E=0$, the terms
$\propto \delta(\Delta E)$ are subdominant, because they only
give a finite contribution to the response, provided that
the scale factor $a$ and the cutoff $k^0$ in~(\ref{propagator:regulated})
are finite and nonzero, while the remaining terms
yield a divergence.

%
%In the ultraviolet ($\Delta E \gg H/2\pi$) 
%this response function behaves as the standard (thermal)
%Maxwell-Boltzmann contribution, ${d{\cal F}(\Delta E)}/{d t}
%   \simeq ({\Delta E}/{2\pi}) {\rm e}^{-2\pi\Delta E/H}$.
%On the other hand, in the infrared ($\Delta E \ll H/2\pi$), 
%it exhibits a more infrared singular behaviour 
%than a thermal distribution, ${d{\cal F}(\Delta E)}/{d t}
%    \simeq (2\pi)^{-2}(H^3/\Delta E^2)$.

However, for a detector sensitive to time separations up to
$\Delta t \leq \Delta_{\rm max}$, ${\cal F}_\delta$
in~(\ref{respone function:rest}) becomes a smeared 
``$\delta$-function,'' 
$d{\cal F}_\delta(\Delta E;\Delta_{\rm max})/dt \propto 
   \sin\big(\Delta E \Delta_{\rm max}/2\big)/(\Delta E/2)$,
such that the detector responds roughly up to the energies
$\Delta E \leq 2 \pi/\Delta_{\rm max}$. Let us now study the question
of finite-time measurements in more detail.

\subsection{Boundary terms through finite-time measurements}
 Strictly speaking, the
propagators~(\ref{Delta:conformally coupled scalar},~\ref{iDelta:4dim:massive},~\ref{iDelta:4dim:massless}) 
describe the dynamics of the scalar field for $t\geq t_0$.
%(a convenient choice is $t_0 = 0$ and $a_0\equiv a(t_0=0) = 1$).
Hence it is natural to consider the Unruh detector, which corresponds to the 
transition amplitude~(\ref{transition amplitude}),
which begins measuring at $t_0=0$ and ends at $t_f=t$. In this case
Eqs.~(\ref{transition probability}) and~(\ref{response function}) 
generalize to 
%
%\begin{equation}
\[
  \frac{d P(E_0\rightarrow E,0,t)}{dt} 
       = h^2 |m_{E_0,E}|^2\; \frac{d{\cal F}(\Delta E,0,t)}{dt}
\qquad (\Delta E = E  - E_0)
\,,
\]
%\label{transition probability:B}
%\end{equation}
%
with
%
%\begin{equation}
\[
\frac{d{\cal F}(\Delta E,0,t)}{dt}  
          = \int_{-t}^{t} d \Delta t 
             \,{\rm e}^{-i\Delta E\Delta t}
                    iG^{<}(t+\Delta t/2,\vec x;t-\Delta t/2,\vec x)
\,,
\]
%\label{response function:B}
%\end{equation}
%
implying that the response function gets modified by the boundary (initial) 
effects. For example, the response function associated with the Hadamard
(conformal) form, $iG_{\rm conf} = H^2/(4\pi^2 y)$, 
for an Unruh detector reads
%
%\begin{equation}
\[
   \frac{d{\cal F}_{\rm conf}(\Delta E,0,t)}{dt} 
            = \frac{\Delta E}{2\pi}\frac{1}{{\rm e}^{2\pi\Delta E/H}-1}
            + \frac{H}{2\pi^2}\sum_{n=1}^\infty n{\rm e}^{-nHt}\,
             \frac{n\cos\big(\Delta Et\big)
                  - (\Delta E/H)\sin\big(\Delta Et\big)}
                  {\big({\Delta E}/{H}\big)^2 + n^2}
\,.
\]
%\[
%   \frac{d{\cal F}_{\rm conf}(\Delta E,0,t)}{dt} 
%            = \frac{\Delta E}{2\pi}\frac{1}{{\rm e}^{2\pi\Delta E/H}-1}
%            + \frac{H}{2\pi^2}\sum_{n=1}^\infty \frac{n}{a^{2n}}
%             \frac{n\cos\Big(\frac{2\Delta E}{H}\ln(a)\Big)
%                  - \frac{\Delta E}{H}\sin\Big(\frac{2\Delta E}{H}\ln(a)\Big)}
%                  {(\Delta E/H)^2 + n^2}
%\,.
%\]
%\label{conformal:t0=0}
%\end{equation}
%
%Note that at late times
%the contributions arising from the finite measurement time
%vanish as ${\rm e}^{-nHt} \; (n=1,2,..)$.
Similar (though more technical) analysis can be performed for other
contributions from the massless scalar propagator~(\ref{iDelta:4dim:massless}).
Quite generically, boundary effects give rise to oscillatory contributions
to the response function of Unruh's detector.
For $\Delta E\sim H$, these terms become unimportant when
$t\gg H^{-1}$.
In the ultraviolet, where $\Delta E\gg H$, the oscillatory contributions
become subdominatnt when $t\gg \Delta E/H^2$ -- much more than a Hubble time.
%This leads to the question, whether the response at first order in perturbation
%theory really describes what would be seen by a physical detector, for more
%details see Ref.~\cite{GarbrechtProkopec:2004}.
Since these effects are however not related to the intrinsic nature
of the quantum fields in de Sitter space, we here do not study them further.

\subsection{Dimensions other than four}
So far, we have calculated the response functions from the scalar propagator,
which is motivated by the practice of defining vacua in de Sitter space
through this quantity.
However, there is a method due to Higuchi~\cite{Higuchi:1986} using a basis of
wave functions as starting point, which we generalize here to $d$ dimensions.
We define
\begin{equation}
\nu=\sqrt{\left(\frac{d-1}2\right)^2-\frac{m_\phi^2+\xi R}{H^2}}\,,
\end{equation}
where the curvature is given by $R=d(d-1)H^2$. The scalar wave equation is
({\it cf.} Eq.~(\ref{nearly minimally coupled light scalar})),
\begin{equation}
\left[\partial_\eta^2+\mathbf{k}^2
           - \frac{\nu^2 - (1/4)}{\eta^2}
 \right]
\varphi_\mathbf k(\eta)=0
\end{equation}
and has the properly normalized positive frequency solution
\begin{equation}
\varphi_\mathbf k(\eta)=\frac 12 (-\pi\eta)^\frac 12
{\rm e}^{\frac\pi 2 \Im[\nu]}H_\nu^{(2)}\left(-|\mathbf k|\eta\right)
\label{phi:solution}
\,.
\end{equation}
The transition probability for the detector in terms of these modes is
then
\begin{eqnarray}
P(\Delta E)\!&=&\!\!
\int\frac{d^{d-1}k}{(2\pi)^{d-1}}\int_{-\infty}^\infty \!\!\!dt \int_{-\infty}^\infty
   \!\!\!dt'
\frac{1}{a(t)a(t')}\varphi^*_\mathbf k(t)\varphi_\mathbf k (t')
{\rm e}^{-i\Delta E(t'-t)}
\\
\!&=&\!\!
\int\frac{d^{d-1}k}{(2\pi)^{d-1}}\int_{-\infty}^\infty  \!\!\!dt \int_{-\infty}^\infty  \!\!\!dt'
\frac{\pi}{4H}{\rm e}^{\pi\Im[\nu]-\frac 32 H(t+t')+i\Delta E(t-t')}
H_\nu^{(2)*}\Big(\frac{|\mathbf k|}{H}{\rm e}^{-Ht}\Big)
H_\nu^{(2)}\Big(\frac{|\mathbf k|}{H}{\rm e}^{-Ht'}\Big).
\nonumber
\end{eqnarray}
From this expression, one obtains for the response function
\begin{equation}
\frac{d{\cal F}_{d}(\Delta E)}{dt}=
\frac{H^{d-3}{\rm e}^{\pi \Delta E/H}}{8\pi^{(d+1)/2} \Gamma\left((d-1)/2\right)}
\left|
\Gamma\left(\frac{(d-1)/2+i\Delta E/H+\nu}{2}\right)
\Gamma\left(\frac{(d-1)/2+i\Delta E/H-\nu}{2}\right)
\right|^2\,,
\label{Higuchi:ddim}
\end{equation}
which reduces for $d=4$ to Higuchi's result~\cite{Higuchi:1986}
\begin{equation}
\frac{d\mathcal{F}(\Delta E)}{dt}=
\frac{H}{4\pi^3}{\rm e}^{-\pi\Delta E/H}
\left|
\Gamma\left(\frac{3/2+i{\Delta E}/{H}+\nu}2\right)
\Gamma\left(\frac{3/2+i\Delta E/H-\nu}2\right)
\right|^2\,.
\label{Higuchi:4dim}
\end{equation}
For $\nu=1/2$ this coincides with our result for the
conformal case~(\ref{response function:conformal}),
and when expanded in ${\tt s}=3/2-\nu$ with the nearly
minimally coupled case~(\ref{respone function:m}).

It is however also interesting to derive some special responses from the scalar
propagator,
which is in $d$ dimensions~\cite{ProkopecPuchwein:2003}
%({\it cf.} Eq.~(\ref{ChernikovTagirov}))
\begin{equation}
%i\Delta_{d}(x,x')
% \rightarrow
iG_{d}(y)
=\frac{\Gamma(\frac{d-1}2+\nu)\Gamma(\frac{d-1}2-\nu)}
{(4\pi)^{d/2}\Gamma\left(\frac{d}{2}\right)}
H^{d-2}\; {}_2F_1\left(\frac{d-1}2+\nu,\frac{d-1}2-\nu,\frac{d}{2};
1-\frac{y}{4}\right)\,.
\end{equation}
In particular, for $d=3$, the response function is exactly
calculable
for arbitrary $\nu$ from the propagator, because we can express the hypergeometric
function in terms of the geodesic
distance $\ell$, using Eq. (9.121.30) of
Ref.~\cite{GradshteynRyzhik:1965}, as
\begin{equation}
_2F_1\left(1+\nu,1-\nu,\frac{3}{2};1-\frac{y}{4}\right)
=\frac{\sin\left[\nu\left(\pi-H\ell\right)\right]}
{\nu\sin\left(H\ell\right)}.
\end{equation}
With $\ell \rightarrow i\Delta t$, 
the response function~(\ref{response function}) is then given by the integral
\begin{equation}
\frac{d{\cal F}_{3}(\Delta E)}{dt}
=\frac{\Gamma(\nu)\Gamma(1-\nu)}{4i\pi^2}H
\int\limits_{-\infty}^{\infty}d\Delta t\, {\rm e}^{-i{\Delta E}\Delta t}\;
\frac{\sin(\pi\nu-i\nu H\Delta t)}{\sinh(H\Delta t)},
\end{equation}
with the poles of the integrand at
$\Delta t_n=i\pi n/H,\,\,n\,\epsilon\, \mathbbm{Z}\backslash \{-1\}$.  Note,
that for odd dimensions, the analytic structure of the propagator is
different than for even dimensions. There are additional poles, but there is
no branch cut (except for special values of $\nu$). 
%Since we consider the Chernikov-Tagirov vacuum,
%there is no singularity at the antipodal
%point, which corresponds to the proper time separation $\Delta t=-i\pi/H$.
We perform the integration by closing the contour in the lower
complex half plane. According to the $\epsilon$-prescription
for the Wightman function $iG^<$, only the poles $n\leq -1$ contribute,
%(the pole $n=-1$ is absent)
and we obtain
\begin{equation}
\frac{d{\cal F}_{3}(\Delta E)}{dt}
=\frac{1}{2}
\frac{\sinh\left(\pi\frac{\Delta E}{H}\right)}
{\cos(\pi\nu)+\cosh\left(\pi\frac{\Delta E}{H}\right)}
\,\frac{1}{{\rm e}^{2\pi{\Delta E}/{H}}-1}\,.
\label{dF/dt:3d}
\end{equation}
Note that~(\ref{dF/dt:3d}) applies not only when $m_\phi<H$, and $\nu$ is real,
but also for $m_\phi > H$, when $\nu$ is imaginary. 
When $m_\phi\gg H, \Delta E$, the response is exponentially
suppressed as $d{\cal F}/dt\propto \exp(-\pi m_\phi/H)$, 
which is a consequence of an exponential suppression of scalar 
particle production in de Sitter space in the limit when $m_\phi\gg H$.

Note that for {\it  no} value of the parameter $\nu$ the
response~(\ref{dF/dt:3d}) agrees with the thermal response function, 
which in three dimensions can be read off from Eq.~(\ref{G:thermal:ddim}),
\begin{eqnarray}
  \frac{d{\cal F}_{3,\tt th}}{dt} 
       = \frac 12 {\rm sign}(\Delta E)\Theta\big((\Delta E)^2 - m_\phi^2\big)
             \,\frac{1}{{\rm e}^{\beta \Delta E} -1}
\,.
\label{response function:2+1:th}
\end{eqnarray}
In particular, for a conformal massless scalar, 
$\nu=1/{2}$, and Eq.~(\ref{dF/dt:3d}) yields
the following `fermionic-like' response function, 
\begin{equation}
\frac{d{\cal F}_{3,\tt conf}(\Delta E)}{dt}
=\frac{1}{2}
 \,\frac{1}{{\rm e}^{2\pi{\Delta E}/{H}}+1}
\,.
\label{dF/dt:3d:conf}
\end{equation}

This disagreement with the thermal case is not a special feature of odd dimensions. {\it E.g.} 
the conformal Green function in $d$ dimensions
\begin{equation}
iG_{{\rm conf},d}=\frac{\Gamma(\frac{d}{2}-1)}{(4\pi^{d/2})}H^{d-2}
y^{1-d/2}
\end{equation}
leads for $d=6$ to the response
\begin{equation}
\frac{d{\cal F}_{6,\tt conf}(\Delta E)}{dt}
=\frac{H^3}{12\pi^2}\left(\frac{\Delta E^3}{H^3}+\frac{\Delta E}{H}\right)
\frac{1}{{\rm e}^{2\pi{\Delta E}/{H}}+1},
\end{equation}
while the flat-space thermal response is
\begin{eqnarray}
  \frac{d{\cal F}_{6,\tt th}}{dt}
      &= \frac{1}{12 \pi^2} {\rm sign}(\Delta E)\Theta\left((\Delta E)^2 - m_\phi^2\right)
             \,\left(\Delta E^2-m^2\right)^{3/2}
\frac{1}{{\rm e}^{\beta \Delta E} -1}
\,.
\label{response function:6:th}
\end{eqnarray}
Generally, for $d>4$, the conformal response consists of the Planck factor
times a polynomial involving different powers of $\Delta E$, therefore
deviating from the thermal response, which involves only
a single power of $\Delta E$.

For $d=2$ conformal and minimally massless coupled case coincide
and we find from~(\ref{Higuchi:ddim})
\begin{equation}
\frac{d{\cal F}_{1+1,{\tt conf}}(\Delta E)}{dt}=\frac{1}{\Delta E}
\frac{1}{{\rm e}^{\beta \Delta E} -1}
\,,\quad(\Delta E\not= 0)\,,
\end{equation}
in agreement with the flat-space thermal response.

Hence, we have found that
an agreement with the thermal response occurs for conformal coupling 
only in $d=2$ and $d=4$, just as for an accelerated
observer in flat space~\cite{GabrielSpindelMassarParentani:1997}.

\section{Detailed balance, response functions and spectra}
\label{Detailed balance, response functions and spectra}

Based on the assumption that the principle of detailed balance holds, 
which states that the  absorbtion rate of the detector ${\cal R}_a$ and
the emission rate ${\cal R}_e$ are equal,
\begin{equation}
  {\cal R}_a(E_0 \rightarrow E) = {\cal R}_e(E \rightarrow E_0)
\,,\qquad
(\,\forall E_0, E\,)
\,,
\label{detailed balance}
\end{equation}
and on the fact that the transition probabilities per unit proper time
are related as follows:
\begin{equation}
   \frac{dP(E_0\rightarrow E)}{dt} = {\rm e}^{-\beta (E-E_0)}
                                     \frac{dP(E\rightarrow E_0)}{dt}
\,,
\label{detailed balance:2}
\end{equation}
or, equivalently, the response function of the detector fulfills,
\begin{equation}
     \frac{d{\cal F}(\Delta E)}{dt}
   = {\rm e}^{-\beta \Delta E}\frac{d{\cal F}(-\Delta E)}{dt}
\qquad (\Delta E = E-E_0)
\,,
\label{detailed balance:dF/dt}
\end{equation}
one can infer that the detector is thermally populated, with the temperature 
given by $T=1/\beta$, as follows
(for a related discussion see Ref.~\cite{SpradlinStromingerVolovich:2001}).
Let us rewrite the principle of detailed balance~(\ref{detailed balance})
as
\begin{equation}
  n(E_0) \frac{dP(E_0\rightarrow E)}{dt}\big(1+n(E)\big)
       = n(E) \frac{dP(E\rightarrow E_0)}{dt} \big(1+n(E_0)\big)
\,,
\label{detailed balance:3}
\end{equation}
where $n(E)$ and $n(E_0)$ denote the occupation numbers of
detector states with energies $E$ and $E_0$, respectively.
%and a chemical potential $\mu$, respectively
%(from the subsequent discussion it follows that in general one should
%allow for a nonvanishing chemical potential). 
%The factors $1+n$ and $1+n_0$ are due to stimulated emission, 
%which characterizes Bose systems. 
%Combining~(\ref{detailed balance:2}) and~(\ref{detailed balance:3}), 
%it follows that $(n^{-1}+1){\rm e}^{-\beta E}$ is a constant,
%indepent on $E$, which we write as ${\rm e}^{\beta\mu}$.
From this, it immediately follows
\begin{equation}
  n(E) = \frac{1}{{\rm e}^{\beta (E-\mu)}-1}
\,,
\label{detector:nEmu}
\end{equation}
such that the  states of the detector are populated according to a
chemical equilibrium at temperature $T = 1/\beta$ and chemical potential $\mu$.

In fact, any  response which can be written as
\begin{equation}
  \frac{d{\cal F}(\Delta E)}{dt}
       = g\; \frac{\Delta E}{2\pi}
         \frac{1}{{\rm e}^{\beta\Delta E} -1}
\,,
\label{detailed balance:confirmed}
\end{equation}
with $g=g(\Delta E)$ being an even function of $\Delta E$, 
fulfills the relation~(\ref{detailed balance:dF/dt}).
We have shown explicitly for different scalar fields in $d=4$
({\it cf.} Eqs.~(\ref{response function:conformal:2}), 
(\ref{respone function:massless}) and~(\ref{respone function:m})), that
they are of the form~(\ref{detailed balance:confirmed}),
\begin{eqnarray}
g_{\rm conf} &=& 1
\nonumber\\
g_{m_\phi = 0} &=& 1+ \Big(\frac{H}{\Delta E}\Big)^2 
             +  2\pi H\delta(\Delta E) \Big[ Ht - \frac 14 + \ln(2)\Big]
\nonumber\\
g_{m_\phi \neq 0} &=& 1+ \Big(\frac{H}{\Delta E}\Big)^2 
             +  2\pi H\delta(\Delta E) \Big[
                                            \frac{1}{2{\tt s}} - 1 + \ln(2)
                                        \Big]
             +  O({\tt s})
\,,
\label{g functions}
\end{eqnarray}
with $\beta_H = 1/T_H = 2\pi/H$. Moreover, the more general expressions
~(\ref{Higuchi:ddim}),~(\ref{Higuchi:4dim}) and~(\ref{dF/dt:3d}) also satisfy
equation~(\ref{detailed balance:dF/dt}).

The relation~(\ref{detailed balance:dF/dt}) can also be viewed
as a consequence of the periodicity of the Green function
$iG^<$ in imaginary proper time, $t\rightarrow t+2\pi i\beta$
\cite{BoussoMaloneyStrominger:2001},
which is in turn a consequence of the same periodicity of the metric 
in Euclidean time
\footnote{For more general de Sitter invariant states, the so called
$\alpha$-vacua, one can show that~(\ref{detailed balance:dF/dt}) does not 
in general hold~\cite{BoussoMaloneyStrominger:2001}. 
However, these states have a different ultraviolet structure than
the Chernikov-Tagirov vacuum~(\ref{ChernikovTagirov}), 
which has the standard Hadamard lightcone
singularity, and hence they are most likely unphysical.}.
An example where the periodicity of the metric however does
not coincide with the Hawking temperature is 
a quasi-de Sitter space considered in Ref.~\cite{BrandenbergerKahn:1982}, 
and hence cannot in general be used as an argument 
for the thermality of a scalar field.

\section{Discussion}

We found the response functions for different scalar fields to differ strongly
in the infrared, where $\Delta E< H$. Moreover, they do not
in general coincide with the response to an equilibrum state in flat space.
A disagreement with the thermal response does  not yet imply that the
detector does not equilibrate with the de Sitter background. In fact, the energy
levels of the detector are thermally populated. Similar deviations
from a Minkowski-space thermal response are also known for accelerated
detectors~\cite{BroutMassarParentaniSpindel:1995,GabrielSpindelMassarParentani:1997}. 
The disagreement of the response functions should be attributed to the fact,
that for the conformally and the minimally coupled scalar field the density
of modes per frequency is different.

However, for fields which are massive or nonconformally
coupled to the metric, the infrared enhancement is a consequence of
the amplification of superhorizon modes leading to cosmological density
perturbations, an effect which is absent in the conformally coupled massless case.
This makes the different fields clearly distinguishable by observables.
The fact that physically distinct situations such as thermally
populated flat space and the different types of de Sitter-invariant vacua
result in the same thermal distribution function for the energy levels of the
detector is due to the insensitivity of this quantity to whether there is
a mixed (thermal) state or a pure (de Sitter-invariant) state and to how
often the detector responds per unit time
\footnote{
An example for the interpretation of the de Sitter invariant states
as thermal can be found in
Ref.~\cite{Higuchi:1986}, where scalar field quantization 
is performed in static coordinates. The mode functions are chosen
to vanish beyond the horizon, where the static coordinates exhibit
a coordinate singularity. Since the horizon distance is singled out,
the mode functions in static coordinates violate spatial homogeneity. 
An Unruh detector placed at the origin of the static vacuum
measures no particles. On the other hand, when the static vacuum is thermally
populated, the response function is the same
as for the de Sitter-invariant vacuum, and it is given 
by~(\ref{Higuchi:4dim}).
Note first that a thermally populated state in static coordinate does
not correspond to a usual thermal equilibrium state, since spatial
homogeneity is broken.
%Moreover, as a consequence of breakdown of spatial homogeneity, 
%two Unruh detectors placed at different points measure different spectra.
%Furthermore, the response of a detector placed away from the origin is not static
%(due to mode mixing). 
In addition, the mode functions in the static
and the de Sitter-invariant vacuum have different
support. Indeed, the static mode functions vanish beyond the de Sitter horizon,
while the de Sitter-invariant mode functions exhibit superhorizon correlations.
%making any comparison between the two vacua questionable. 
%With these remarks in mind, we argue that the relevance of the static vacuum 
%is questionable in cosmological applications.
}.

Finally, we want to point out that de Sitter invariance of the
stress-energy tensor of the conformal vacuum implies that 
the stress-energy tensor must be proportional to the metric, 
such that $p_{\rm conf} = -\rho_{\rm conf}$~\cite{Brandenberger:1983}.
This is inconsistent with a thermal equation of state and leads us to
the investigation, whether the energy density is captured at higher
orders in perturbation theory~\cite{GarbrechtProkopec:2004}, allowing for 
the pair creation operators
of the scalar field Hamiltonian, which are not captured by the first order
perturbation expansion used here. While we argued
in this paper that the thermal state of the detector does not contain
the full information about the quantum field, it is yet remarkable, that
at first order in perturbation theory the detector is insensitive to the
(regulated) stress-energy tensor, which is clearly nonthermal.

\section*{Acknowledgements} 

 We would like to thank
Michael G. Schmidt and Richard P. Woodard
for critical reading of the manuscript and for many suggestions,
and  Robert H. Brandenberger,  Atsushi Higuchi, Claus Kiefer, Klaus Kirsten
and Bharat Ratra for useful correspondence. 
T.P. thanks Kiril Krasnov for discussions. T.P. acknowledges 
a generous financial support of the Albert Einstein Institute in Golm.

%
%%%%%%%%%%%%%%%%%%%%%%%%%%%%%%%%%%%%%%%%%%%%%%%%%%%%%%%%%%%%%%%%%%%%%%%%%%%%%%%
%    REFERENCES
%%%%%%%%%%%%%%%%%%%%%%%%%%%%%%%%%%%%%%%%%%%%%%%%%%%%%%%%%%%%%%%%%%%%%%%%%%%%%%%
%


\begin{thebibliography}{99}
\bibliographystyle{unsrt}



%\cite{Unruh:db}
\bibitem{Unruh:1976}
W.~G.~Unruh,
``Notes On Black Hole Evaporation,''
Phys.\ Rev.\ D {\bf 14} (1976) 870.
%%CITATION = PHRVA,D14,870;%%

\bibitem{BirrellDavies:1984}
%\cite{Birrell:ix}
%\bibitem{Birrell:ix}
N.~D.~Birrell and P.~C.~W.~Davies,
``Quantum Fields In Curved Space,''
Cambridge University Press (1984).
%\href{http://www.slac.stanford.edu/spires/find/hep/www?irn=998621}{SPIRES entry}

%\cite{Gibbons:mu}
\bibitem{GibbonsHawking:1977}
G.~W.~Gibbons and S.~W.~Hawking,
``Cosmological Event Horizons, Thermodynamics, And Particle Creation,''
Phys.\ Rev.\ D {\bf 15} (1977) 2738.
%%CITATION = PHRVA,D15,2738;%%

%\cite{Spradlin:2001pw}
\bibitem{SpradlinStromingerVolovich:2001}
M.~Spradlin, A.~Strominger and A.~Volovich,
``Les Houches lectures on de Sitter space,''
published in ``Les Houches 2001, Gravity, gauge theories and strings'' 423-453 
[arXiv:hep-th/0110007].
%%CITATION = HEP-TH 0110007;%%

%\cite{Bousso:2001mw}
\bibitem{BoussoMaloneyStrominger:2001}
R.~Bousso, A.~Maloney and A.~Strominger,
 ``Conformal vacua and entropy in de Sitter space,''
Phys.\ Rev.\ D {\bf 65} (2002) 104039
[arXiv:hep-th/0112218].
%%CITATION = HEP-TH 0112218;%%

%\cite{Higuchi:1986}
\bibitem{Higuchi:1986}
A.~Higuchi,
``Quantization Of Scalar And Vector Fields Inside The Cosmological Event
Horizon And Its Application To Hawking Effect,''
Class.\ Quant.\ Grav.\  {\bf 4} (1987) 721.
%%CITATION = CQGRD,4,721;%%

%\cite{Chernikov:zm}
\bibitem{ChernikovTagirov:1968}
N.~A.~Chernikov and E.~A.~Tagirov,
``Quantum Theory Of Scalar Fields In De Sitter Space-Time,''
Annales Poincare Phys.\ Theor.\ A {\bf 9} (1968) 109.
%%CITATION = AHPAA,A9,109;%%

\bibitem{BunchDavies:1978}
T.~S.~Bunch and P.~C.~Davies,
``Quantum Field Theory In De Sitter Space: 
 Renormalization By Point Splitting,''
Proc.\ Roy.\ Soc.\ Lond.\ A {\bf 360} (1978) 117.
%%CITATION = PRSLA,A360,117;%%

\bibitem{Mottola:1984}
E.~Mottola,
``Particle Creation In De Sitter Space,''
Phys.\ Rev.\ D {\bf 31} (1985) 754.
%%CITATION = PHRVA,D31,754;%%
%\cite{Allen:ux}

\bibitem{Allen:1985}
Bruce Allen,
``Vacuum States In De Sitter Space,''
Phys.\ Rev.\ D {\bf 32} (1985) 3136.
%%CITATION = PHRVA,D32,3136;%%

\bibitem{AllenFolacci:1987}
B.~Allen and A.~Folacci,
``The Massless Minimally Coupled Scalar Field In De Sitter Space,''
Phys.\ Rev.\ D {\bf 35} (1987) 3771.
%%CITATION = PHRVA,D35,3771;%%

\bibitem{LeBellac:1996} M.~Le~Bellac, ``Thermal Field Theory,''
Cambridge University Press (1996).

%\cite{Prokopec:2003tm}
\bibitem{ProkopecPuchwein:2003}
T.~Prokopec and E.~Puchwein,
``Photon mass generation during inflation: de Sitter invariant case,''
JCAP {\bf 0404} (2004) 007
[arXiv:astro-ph/0312274].
%%CITATION = ASTRO-PH 0312274;%%

\bibitem{GradshteynRyzhik:1965}
Izrail Solomonovich Gradshteyn, Iosif Moiseevich Ryzhik,
{\it Table of integrals, series, and products}, 4th edition, 
Academic Press, New York (1965).  

\bibitem{OnemliWoodard:2004}
V.~K.~Onemli and R.~P.~Woodard,
``Quantum effects can render w < -1 on cosmological scales,''
arXiv:gr-qc/0406098.
%%CITATION = GR-QC 0406098;%%

%\cite{Prokopec:2003iu}
\bibitem{ProkopecWoodard:2003}
T.~Prokopec and R.~P.~Woodard,
``Dynamics of super-horizon photons during inflation with vacuum
polarization,''
Annals Phys.\  {\bf 312} (2004) 1
[arXiv:gr-qc/0310056].
%%CITATION = GR-QC 0310056;%%


\bibitem{KirstenGarriga:1993}
K.~Kirsten and J.~Garriga,
``Massless Minimally Coupled Fields In De Sitter Space: O(4) Symmetric States
Versus De Sitter Invariant Vacuum,''
Phys.\ Rev.\ D {\bf 48} (1993) 567
[arXiv:gr-qc/9305013].
%%CITATION = GR-QC 9305013;%%

\bibitem{TsamisWoodard:1993}
N.~C.~Tsamis and R.~P.~Woodard,
%``The Physical basis for infrared divergences in inflationary quantum
%gravity,''
Class.\ Quant.\ Grav.\  {\bf 11} (1994) 2969.
%%CITATION = CQGRD,11,2969;%%

\bibitem{GabrielSpindelMassarParentani:1997}
C.~Gabriel, P.~Spindel, S.~Massar and R.~Parentani,
``Interacting charged particles in an electric field and the Unruh  effect,''
Phys.\ Rev.\ D {\bf 57} (1998) 6496
[arXiv:hep-th/9706030].
%%CITATION = HEP-TH 9706030;%%

%\cite{Brandenberger:xi}
\bibitem{BrandenbergerKahn:1982}
R.~H.~Brandenberger and R.~Kahn,
``Hawking Radiation In An Inflationary Universe,''
Phys.\ Lett.\ B {\bf 119} (1982) 75.
%%CITATION = PHLTA,B119,75;%%

\bibitem{BroutMassarParentaniSpindel:1995}
R.~Brout, S.~Massar, R.~Parentani and P.~Spindel,
``A Primer for black hole quantum physics,''
Phys.\ Rept.\  {\bf 260} (1995) 329.
%%CITATION = PRPLC,260,329;%%


%\cite{Brandenberger:1983gy}
\bibitem{Brandenberger:1983}
R.~H.~Brandenberger,
``An Alternate Derivation Of Radiation In An Inflationary Universe,''
Phys.\ Lett.\ B {\bf 129} (1983) 397.
%%CITATION = PHLTA,B129,397;%%


\bibitem{GarbrechtProkopec:2004}
B.~Garbrecht and T.~Prokopec,
``Energy density in expanding universes as seen by Unruh's detector,''
arXiv:gr-qc/0406114.

\end{thebibliography}
\end{document}